\documentclass[oldversion]{aa} 
\usepackage{graphicx}
\usepackage{txfonts}
\usepackage{natbib}
\def\wx{WX\,LMi}
\def\Na{Na}
\def\na1{Na{\sc I}}
\def\crat{s$^{-1}$}
\def\porb{P_{\rm orb}}
\def\hal{H$\alpha$}
\def\msun{M$_\odot$}

\def\rwd{R$_{\rm wd}$}

\def\ftot{erg\,cm$^{-2}$\,s$^{-1}$}
\def\mrat{M$_\odot$\,yr$^{-1}$}
\def\snd{$2^{\rm nd}$}
\def\third{$3^{\rm rd}$}
\def\fourth{$4^{\rm th}$}

\def\fispe{\phi}
\def\teff{T_{\rm eff}}

\begin{document}

   \title{An in-depth study of the pre-polar candidate \wx}

   \author{J. Vogel\inst{1},
     A.D. Schwope\inst{1}
     and B.T.G{\"a}nsicke\inst{2}
   }

   \institute{Astrophysikalisches Institut Potsdam,
                An der Sternwarte 16, D-14482 Potsdam\\
              \email{jvogel@aip.de}
              \and
              Department of Physics, University of Warwick, Coventry, CV4 7AL,
                UK}

   \date{Received ; accepted }

\abstract{Optical photometry, spectroscopy and XMM-Newton ultraviolet and
  X-ray observations with full phase coverage are used for an in-depth study
  of \wx, a system formerly termed as a low-accretion rate polar. We find a
  constant low mass accretion rate, $\dot{M} \sim 1.5 \times 10^{-13}$\mrat, 
  a peculiar accretion geometry with one spot not being accessible via
  Roche-lobe overflow, a low temperature of the white dwarf, $\teff <
  8000$\,K and the secondary likely being Roche-lobe
  underfilling. All this lends further support to the changed view on \wx\ and related
  systems as detached binaries, i.e.~magnetic post-common envelope binaries
  without significant Roche-lobe overflow in the past. 
  The transfer rate determined here is compatible with
  accretion from a stellar wind. We use cyclotron spectroscopy to determine
  the accretion geometry and to constrain the plasma temperatures. Both,
  cyclotron spectroscopy and X-ray plasma diagnostics reveal low plasma
  temperatures below 3\,keV on both accretion spots. For the low $\dot{m}$,
  high $B$ plasma at the accretion spots in \wx, cyclotron cooling is
  dominating thermal plasma radiation in the optical.
  Optical spectroscopy and X-ray timing reveal atmospheric, chromospheric and
  coronal activity at the saturation level on the dM4.5 secondary star.
}
\maketitle

\section{Introduction}
Polars are magnetic cataclysmic binaries consisting of a late-type 
main-sequence star and a strongly magnetic white dwarf locked in synchronous 
rotation (see \cite{1995cvs..book.....W} for a comprehensive survey of 
cataclysmic variable stars). In normal polars accretion happens via Roche-lobe overflow and 
accretion streams towards the magnetic poles where the accretion energy 
is released mainly as X-ray thermal radiation, optical cyclotron radiation and
a prominent soft X-ray component. The soft component makes them prominent
sources in the soft X-ray sky and they were found numerously in the ROSAT
all-sky survey. The detection bias is large and the true space  density highly
uncertain. Recently a few systems with very low accretion rate (a factor 100
-- 1000 below the canonical values for Roche-lobe  overflow) were uncovered in
optical spectroscopic surveys (HQS: \citealt{1995A&AS..111..195H}, 
SDSS: \citealt{2000AJ....120.1579Y}). They were identified on the basis
of highly peculiar optical spectra with  broad cyclotron harmonics in emission
mimicing quasar emission lines. They could form an important part of the
population of close interacting binaries, being either normal CVs in extended
low states or pre-CVs on their track towards Roche-lobe
overflow. Recently, pushed forward by the work by \cite{2005ApJ...630.1037S}
evidence grows to regard these systems as pre-CVs.
They might serve as tracers to uncover the unbiased sample of
magnetic CVs.  

\wx\ (HS 1023+3900) was discovered by \cite{1999A&A...343..157R} from the Hamburg Quasar
Survey. From optical and spectroscopic follow-up observations they  
determined a period of about 167 minutes and two accretion spots with a field
strength of 60 MG and 68 MG, respectively.  
The accretion rate was found to be $\dot{M} < 3 \times 10^{-13}$
M$_{\odot}$/yr, i.e.~orders of magnitudes below the normally observed  
accretion rate in polars. A spectral type of dM4.5 gave a good spectral fit
for the secondary and was used to deduce a distance of $140 \pm 50$
pc. \cite{2001A&A...374..189S} used optical photometry, performed with the
Potsdam 70 cm telescope in {\it UBVRI}, to determine a stable long-term ephemeris
based on the timings of optical maxima from the primary accretion spot. 
During the whole monitoring campaign covering half a year the system was
in a similar low state of accretion with the one exception of a flare which
was located on the active secondary.
From the colors at the orbital phase when both
accretion spots are invisible the spectral type of the secondary was likewise
determined to be M4.5. Using the relative magnitude in the I band at
photometric minimum yields a distance of $\sim$ 100 pc.  

\wx\ as well as the similar objects termed LARPs (Low Accretion Rate Polars,
\cite{2002ASPC..261..102S}) was not discovered
in the ROSAT All Sky Survey despite its relatively short distance. 
\wx\ was in the off-axis field of two X-ray ROSAT PSPC pointings (10 and 13
ksec) and discovered at a rate of 0.004 cts/sec. The very low number of
photons did not allow to extract a proper X-ray spectrum and to determine the
origin of the X-rays, being either accretion-induced or from the active corona
of the secondary. We thus performed new X-ray observations with XMM-Newton in
order to study magnetic accretion in an extreme combination of the 
parameters, which control the process of energy release, i.e. at very low
specific accretion rate $\dot{m}$ and at high magnetic field. 
The X-ray observations of \wx\ were accompanied by near-ultraviolet
observations with the OM onboard XMM-Newton. It was used as a bolometer in
order to determine the white dwarf photospheric temperature. This should help 
to decide if the LARPs are normal polars in an 
occasional state of very low accretion or pre-cataclysmic binaries that 
never have been accreting. 
Our newer X-ray and UV observations are analysed together with phase-resolved
optical spectroscopic observations performed already back in 1999. Since the
overall brightness of \wx\ does not change, the two data sets can be combined
without problem and thus allow a broader picture of the relevant spectral
components. 

The paper is organized as follows. In Section~\ref{data} the new optical and
X-ray observations and the reduction steps are described. In Sect.~3 we begin
the analysis by an in-depth study of the spectral features from the secondary
which constrains the spectral types, the distance, its activity and the binary
parameters. In Sect.~3.2 we describe our modeling of the phase-resolved
cyclotron spectra, which constrains the locations of the accretion spots and
the temperatures in the accretion plasmas. Sect.~3.3 describes our attempts to
understand the OM-filter observations and Sect.~3.4 finally gives an analysis
of the X-ray spectrum. 
\begin{table}[t]
\caption{Log of spectroscopic observations with the Calar Alto 3.5m telescope} 
\label{table:2}                        
\renewcommand{\footnoterule}{}
\begin{tabular}{cccccc}       
\hline\hline
Date         &  Instrument & Wavelength  & Resolution & No. of\\
(Y/M/D)      &             & range [\AA] &   [\AA]    & spectra$^b$\\
\hline
1999/3/9     &  TWIN       & 3800-9900   & 6/4$^a$ & 47\\
1999/3/10    &  TWIN       & 3800-9900   & 6/4$^a$ & 56\\
1999/3/18    &  MOSCA      & 3380-8410   & 13      & 29\\
\hline
\end{tabular}\\
$^a$ for the blue/red channel \\
$^b$ integration time 300 seconds each 
\end{table}
\section{New optical and X-ray observations}
\label{data}
\subsection{Optical photometry -- updated ephemeris}
\label{ephemeris}
\cite{2001A&A...374..189S} fixed the ephemeris of the system using photometric
data covering a period of six month. The primary maxima of the V  
band data pinned down the epoch of zero phase and the variability of the light
curves in VRI was used to establish the period. We  
performed additional photometric observation with the Potsdam 70cm telescope
in V band on April 28 2004 and March 21 2005. 
These observations revealed the system to be at the same brightness as during
our monitoring observations in 1999 and also showed the same variability pattern.
A linear regression to all observed primary V band maxima between 1999 and 2005 
yield a period of 0.11592364(43) days. Our spectroscopic observations from March
1999 determine the inferior conjunction of the secondary. The epoch of the 
blue-to-red zero crossing of the \na1-lines (see Sect. \ref{zeropoint}) was used
as the spectroscopic zero point. This yields the updated ephemeris
\begin{equation}
\mbox{BJD} = 2451247.50425(27) + E \times 0.11592364(43)
\end{equation}
where the number in brackets represent the uncertainty. The difference between
spectroscopic and photometric phase zero is 0.168 phase units. The phase 
used in this paper refers to this spectroscopic ephemeris.
\begin{figure}[t]
\resizebox{\hsize}{!}{\includegraphics[clip=]{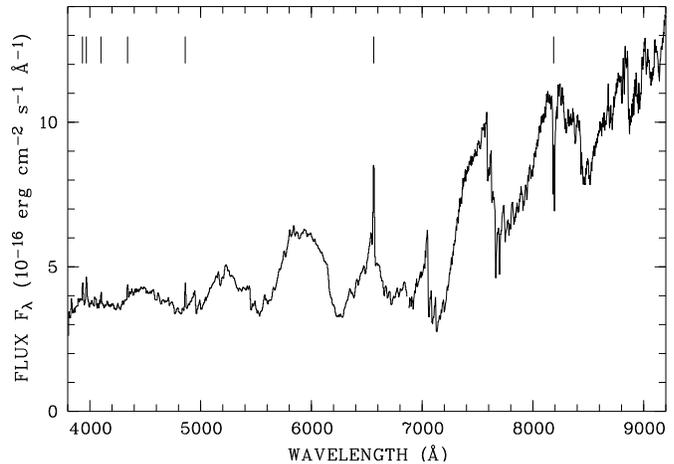}}
\caption{Mean orbital spectrum of \wx\ obtained March/April, 1999, with the
  TWIN and MOSCA spectrographs at the Calar Alto 3.5m telescope. The Ca H+K,
  the Balmer emission lines and Na absorption lines are indicated by vertical
  ticks.
} 
\label{f:means}
\end{figure}
\subsection{Optical spectroscopy}
\label{optspec}
\wx\ was observed with the Calar Alto 3.5m telescope at two occasions in
March 1999. The double-beam spectrograph TWIN was used for the first run in
March 9--11, the multi-object spectrograph MOSCA for the second run on March
18/19.  
Low-resolution gratings with reciprocal dispersions of 144\,\AA/mm and
108\,\AA/mm were used for the TWIN observations resulting in a spectral
coverage of 3800--6900\,\AA\ and 6700--9900\,\AA\ in the blue and red channels
at a spectral resolution of about 6 and 4\,\AA, respectively. 
A series of 47 spectra with exposure time of 5 minutes each were obtained in
the night March 9/10 (BJD 2451247.484406 -- 2451247.671117), another series of
56 spectra with the same exposure time in the following night (BJD
2451248.375258 -- 2451248.596263). All binary phases were thus covered by
spectroscopic observations.  

Spectrophotometric standards were observed in the same nights and used for calibration of the
spectral response. However, the weather conditions were rather poor during the
TWIN observations with transparency variations by 100\%. Hence, we could not
use the standard star observations to put the spectra of \wx\ on an absolute
scale, we could just determine the spectral slope.
The nearby star GSC2-N23233202253 ($\alpha$ (2000):
10:26:27.5, $\delta$ (2000): +38:45:03, distance to \wx: $86\arcsec$) was
used as comparison star and also placed on the 
spectrograph's slit. We tried to use the variable signal of
this star for a photometric calibration of the spectra of our target. 
This attempt turned out not to be successful since the photometric variations
showed color-terms, likely to be different between the
target and the comparison star (likely/possible reasons for that being a
slight misalignement of the spectrograph slit, observation not being performed
at the parallactic angle, the different spectral slopes of the target and the
comparison).
 
Instead we used our photometric observations at the 70cm telescope to
calibrate our spectra (assuming no overall change of brightness and
color). This approach was also not found to be free of problems since the
wavelength ranges of the spectrograph channels were too narrow to cover more
than one (red) or two (blue) filters completely which made a correction of
the color terms rather uncertain.
 
He/Ar arc lamp spectra were observed for calibrating the dispersion. 
Flexure of the spectrograph resulted in considerable shifts of
line positions on the CCD of up to 0.6 pixels. These were measured in each
spectrum individually by correlating the night-sky spectrum with a template. 
These time-dependent shifts were then approximated by a low-order polynomial
and the dispersion relation was shifted per spectrum according to those fits.
The late-type stars Gl213, Gl268, and Gl273 were observed as possible template 
stars for the main-sequence secondary in \wx\ through wide and narrow slits in
the night of March 10.

We followed a similar observational strategy for the MOSCA observations by
placing the same comparison star on the spectrograph's slit and obtained 29
exposures of 5 minutes each in the night March 18/19 between UT 22:45 and 1:41
(BJD 451256.454120 -- 2451256.576432), i.e. we covered a bit less than one
orbital cycle of the 167 min binary. The B500 grism was used as a disperser
resulting in a 
spectral coverage 3380 -- 8410\,\AA\ with about 13\,\AA\ (FWHM) resolution. 
Similar to the TWIN observations, small remaining shifts of the positions of
the night sky lines in each spectrum were used to improve the dispersion
relation which was originally based on HgNeAr arc lamp spectra.
 
Also the MOSCA observations suffered from large transparency variations by almost
100\%. This time these variations could rather successfully corrected for with
the help of the 
comparison star observed through the same slit. A further complication arose
from the fact that no spectrophotometric standard stars could be observed
during that night. A photometric calibration of the MOSCA spectra was
achieved by assuming that the orbital mean spectrum of \wx\ was unchanged
between the TWIN and the MOSCA observations, an assumption which is justified
by our long-term photometry \citep{2001A&A...374..189S}. Division of the
uncalibrated mean MOSCA spectrum by the calibrated mean TWIN spectrum yielded
an instrumental response curve for the MOSCA observations.   

\subsection{XMM-Newton X-ray and ultraviolet observations}
\label{xmm}
We performed ultraviolet and X-ray observations of \wx\ with  XMM-Newton
in April 2004 for about 30 ksec, covering 3 orbital cycles. Calibrated photon
event tables for all detectors were computed with SAS version 6.5. Since the
spectrum turned out to be very soft, the SAS task {\tt epreject} was used
to extend the usable energy range down to 0.12 keV. Barycentric correction was
applied using {\tt barycen}. The spectra were extracted from the event tables 
with SAS tasks {\tt evselect} and {\tt especget}. The light curve was obtained 
running the {\tt edetect\_chain} pipeline on time selected event tables containing 
all events ocurred in a given orbital phase interval.

The EPIC PN was operated in full frame imaging mode, yielding $\sim$ 800 counts 
from the source. Both EPIC MOS cameras were operated in partial window mode, 
yielding both $\sim$ 200 counts. The  observing time with the OM was spent 
in nearly equal shares with the U and UVW1 filters centered on 3440\,\AA\ 
and 2910\,\AA, respectively. The detector was used in fast mode for the full
observation. The  mean countrates in the U and UVW1 filters were 2.12\,\crat
and 0.75\,\crat, respectively. 

\begin{table}[t]
\caption{Parameters for the secondary in \wx
\label{t:m2par}
}
\begin{tabular}{cccccccc}
\hline\hline
Sp & 
$\log(R/R_\odot)^a$ & 
$M/M_\odot^b$ & 
$M_K^c$ & 
$S_K^d$ & 
$F_{\rm TiO}^e$ & 
$d(S_K)$ & 
$d^f({\rm TiO}) $ \\
\hline
4  &-0.613& 0.220& 7.53& 4.37& 3.08& 103& 130\\
4.5&-0.676& 0.179& 7.95& 4.46& 2.62&  85& 104\\
5  &-0.740& 0.147& 8.37& 4.56& 2.15&  70&  82\\
\hline
\end{tabular}
$^a$ according to \cite{1999A&A...348..524B}\\
$^b$ according to \cite{2000A&A...364..217D}\\
$^c$ according to \cite{2000A&A...364..217D}\\
$^d$ according to \cite{1999A&A...348..524B}\\
$^e$ according to \cite{1999A&A...348..524B} in units of $10^5$\,erg\,cm$^{-1}$\,\crat\,\AA$^{-1}$\\
$^f$ using an observed flux deficit $F_{7165} = 5.5 \times
10^{-16}$\,erg\,cm$^{-1}$\,\crat\,\AA$^{-1}$ (uncertainty of 30\%)
\end{table}

\section{Analysis}
\label{analyse}

Fig.~\ref{f:means} shows the mean orbital spectrum of \wx\ obtained with the
TWIN  (longward 6800\,\AA) and MOSCA spectrographs. It shows the same features
as described by \cite{1999A&A...343..157R}, i.e. an M dwarf dominating the red
spectral range, a white dwarf which is responsible for the spectral upturn in
the blue range and several pronounced cyclotron lines, the most prominent one 
at 5950\,\AA. The higher spectral resolution and the partly better signal
provided by the observations presented here allows individual spectral
features to be resolved and studied. The most prominent ones are the H-Balmer
emission lines and the \na1 absorption lines at 8183/94\,\AA. We
therefore begin our analysis with an optical spectroscopic study focusing on
the secondary star and the optical cyclotron spectrum.

\subsection{The secondary in \wx\ -- optical spectroscopy}
\subsubsection{Spectral type of the secondary}
\label{sptype}
In order to determine the spectral type of the secondary we make use of
the TWIN spectra obtained in the red channel.
The spectral type is not determined straightforwardly due 
to the wavelength- and phase-dependent background component from the white
dwarf (photospheric and beamed cyclotron radiation from two accretion spots).
The remaining calibration uncertainties, as described above, gave rise to some
additional complication.\\ 
We firstly shifted all spectra to radial velocity zero using the sine fit to
the \Na~radial velocity curve (see below). We then selected only those
spectra which were not too heavily affected by atmospheric absorption,
i.e.~those with the best signal-to-noise ratio, and calculated an average  
spectrum.\\
We adapted a non-magnetic model atmosphere of the white dwarf 
to the average spectrum in a wavelength region free of any cyclotron emission
and subtracted the same scaled model spectrum from all selected observed
spectra. This left us with phase-dependent spectra containing only the
secondary and the cyclotron lines. \\ 
In order to constrain the spectral type of the secondary 
we decided to use the narrow-band spectral indices TiO5 and
VO-a (\cite{2002AJ....123.2828C}, \cite{1999ApJ...519..802K}). 
According to those papers these indices correlate best with the spectral
type. We also tried the PC3 index \citep{1999AJ....118.2466M} although we did regard the
results less reliable since the two continuum side bands have a relatively
large separation and the index is therefore much more sensitive to variations
of the underlying cyclotron radiation.\\
We determined the indices for all spectra, eliminated those indices where the
spectral features were found superposed on a cyclotron line and finally
calculated average indices. The TiO5 index thus determined was $0.30\pm0.05$,
the VO-a index $1.99\pm0.03$, and the PC3 index to $1.21\pm0.05$.
The given errors reflect the statistical
uncertainty and the uncertainty which originates from the subtraction of the
white dwarf model spectrum. The measured indices resulted in spectral types
of M5.0$\pm0.5$, M4.5$\pm0.7$, and M4.5$\pm1$. We finally adopted a spectral
type of M4.5 with an uncertainty of half a subclass. This new
determination of the spectral type based on spectral data with sufficient
resolution is in good agreement with the estimates of \cite{1999A&A...343..157R}
and \cite{2001A&A...374..189S} based on spectroscopy with much lower resolution 
and broad-band optical photometry, respectively.\\
\cite{1998A&A...339..518B} have shown that the color $I-K$ can be used as
tracer of the spectral type. A parameterization of the Sp-$(I-K)$ relation 
was given as a third-order polynomial. Due to the high field strenght in \wx\ the $K$-Band 
is likely uncontaminated by cyclotron radiation (see Fig.~\ref{sed} for the wavelength of 
the cyclotron fundamental for both poles), thus allowing to determine a spectral
type of M4 for $I = 14.83$ \citep{2001A&A...374..189S} and $K = 12.49$ 
(2MASS database) in accordance with the above value. \cite{1998A&A...339..518B}
have also shown that most CV secondaries with $\porb$ between 2 and 5 hours
are cooler than expected for Roche-lobe filling ZAMS stars. 
Nuclear evolution prior to mass transfer and lack of thermal equilibrium 
due to mass loss were mentioned as possible causes of this discrepancy. 
We will argue below that both options likely do not apply here
suggesting the presence of a Roche-lobe underfilling main-sequence secondary
in \wx. 

\subsubsection{The distance to \wx}
\label{dist}
The composite nature of the emission from cataclysmic variables, particular in
the optical, makes it difficult to perform a straightforward distance
measurement. We do not make use of Roche lobe geometry, since the secondary 
is possibly somewhat underfilling. The simple use of the distance modulus also
doesn't seem to be appropriate, since the secondary is clearly deformed as
evidenced from ellipsoidal modulations (Fig.~\ref{f:halpha}, bottom panel). 
We assume that the secondary in \wx\
is a ZAMS star with solar metallicity and make use of the surface brightness -
spectral type and $F_{\rm TiO}$ - spectral type relations compiled by
\cite{1999ASPC..157..283B}. We use radii from the compilations of \cite{1999A&A...348..524B}. 
Absolute magnitudes and masses for the possible spectral types
are derived from \cite{2000A&A...364..217D}. The results are summarized in
Table~\ref{t:m2par}. 

The distance via the surface brightness $S_K$ assumes that the 2MASS $K$-band
magnitude is solely due to the secondary star. The measured flux deficit in
the TiO feature was corrected for the contribution of the white dwarf at this
wavelength. 
For the most likely spectral type, M4.5, our distance estimates give  85
and 100\,pc, respectively, with a relatively large spread of about 20\,pc. 
We regard the derived distance via the $S_K$
relation somewhat more reliable, since the photometric data are not affected
by the calibration problems of our spectroscopy. If the secondary would behave
like most of the CV secondaries at the given $\porb$, the radii would be
larger than given in Tab.~\ref{t:m2par} and the distance would increase. 
We thus use in the following a distance of 100\,pc to \wx.
\begin{figure}[t]
\resizebox{0.49\hsize}{!}{\includegraphics[bb=101 233 511
  637,clip=]{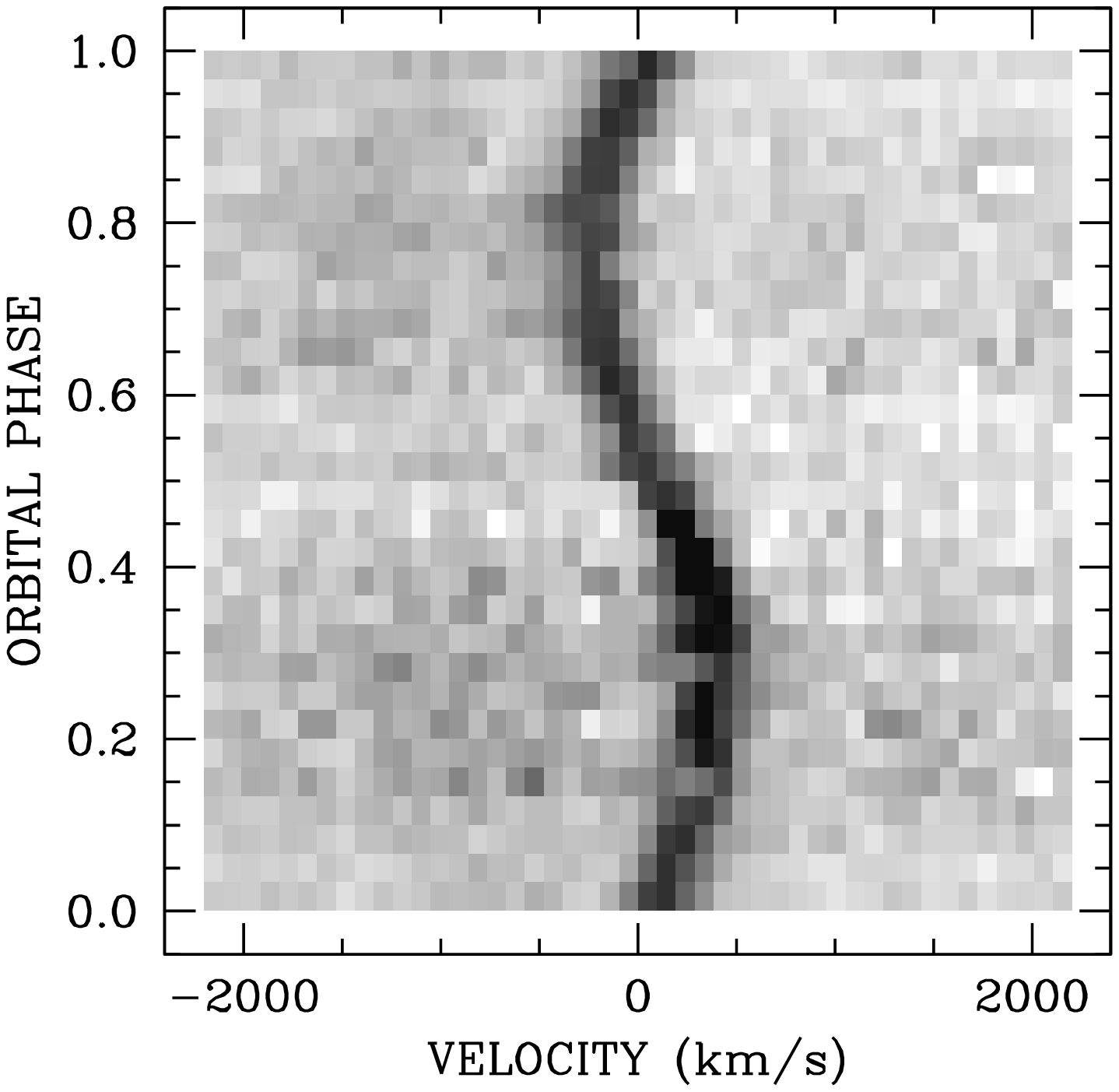}}
\hfill
\resizebox{0.49\hsize}{!}{\includegraphics[bb=101 233 511
  637,clip=]{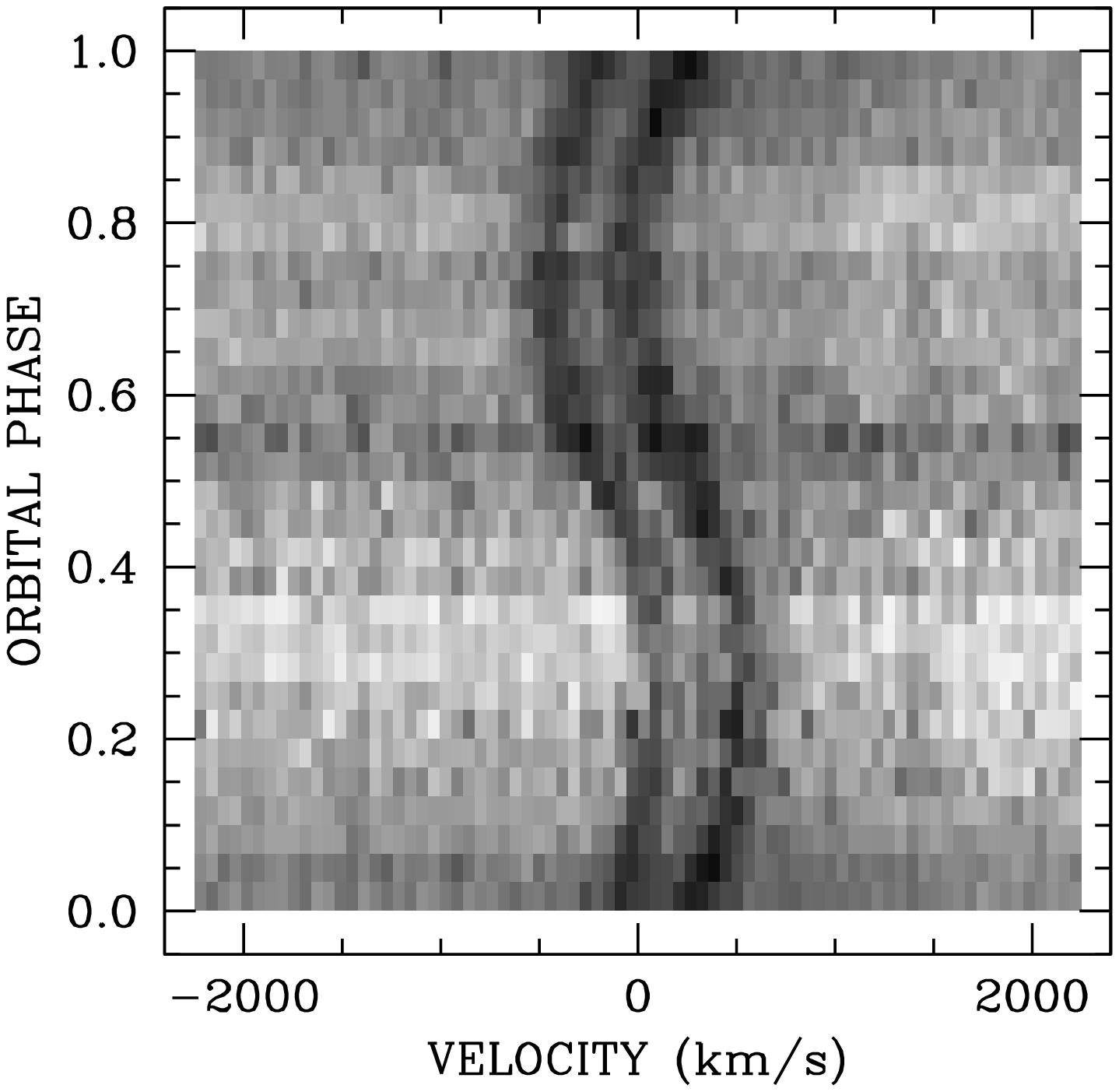}} 
\caption{TWIN spectroscopy of \wx. Shown are cutouts of the phase-averaged
  spectra arranged as an apparent trailed spectrogram centered on the \hal\
  emission line and the \na1 absorption lines. Phase zero refers to inferior
  conjuntion of the secondary, Doppler shifts were transformed to velocity.
}
\label{f:msl}
\end{figure}

\begin{figure}[t]
\resizebox{\hsize}{!}{\includegraphics[clip=]{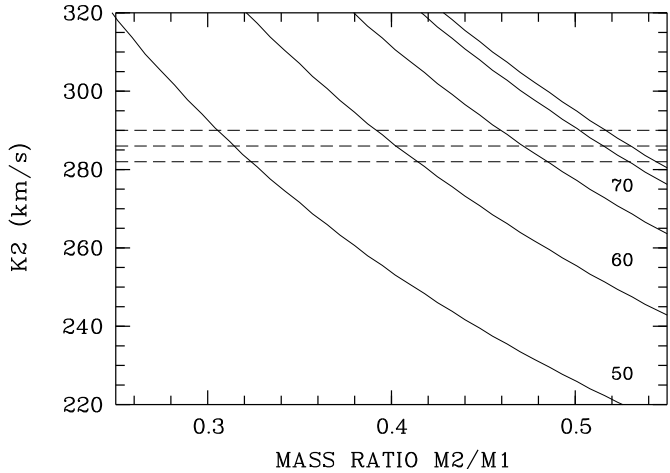}}
\caption{Radial velocity of the secondary as a function of the mass ratio and
  the orbital inclination (between 50\degr\ and 90\degr\ in steps of
  10\degr\,). The horizontal dashed lines mark the observed radial velocity
  with  1$\sigma$ uncertainties.}
\label{f:qK2}
\end{figure}

\subsubsection{Emission and absorption line radial velocity variations}
\label{zeropoint}
Our final reduced spectra were phase-averaged and arranged as an apparent
trailed spectrogram using our improved orbital period and the phase of
inferior conjunction of the secondary as determined 
below. A cutout of the spectra centered on the H$\alpha$ emission lines and
the \na1\ absorption lines is shown in Fig.~\ref{f:msl}. Wavelength was
transformed to radial velocity in the diagrams. The lines from both species
are clearly detected through the complete orbital cycle.

The \na1-doublet is clearly resolved in the TWIN spectra. Radial velocity
measurements were performed by fitting double Gaussians with fixed wavelength
separation to the observed line profiles. Those attempts were not always
successful due to the low signal to noise ratio in several of the spectra. For
the determination of the radial velocity curve we neglected spectra below a
certain S/N level and were thus left with a total of 65 spectra.

Despite the lack of sufficient resolution the H$\alpha$ lines of \wx\ seem to 
consist of just one line. Contrary to polars in their high accretion states, 
which show even at our spectral resolution more complex line profiles.
H$\alpha$ line positions in the individual spectra were thus
determined by fitting single Gaussians. The H$\alpha$ lines were detected also 
in all MOSCA spectra and line positions were determined by fitting Gaussians 
in 26 spectra.

The radial velocity curves of \na1\ and of H$\alpha$ at both occasions could
be successfully fitted with sine curves. A summary of the radial velocity
measurments is given in Table~\ref{t:vrad}. The epoch of the blue-to-red zero
crossing of the \na1-lines at BJD = 2451247.50425(27) determines inferior
conjunction of the secondary and spectroscopic phase zero
throughout this paper. The number in parentheses gives the uncertainty in the
last digits of the ephemeris zero point. 
\begin{table}[t]
\caption{Radial velocity measurements of \wx\ in 1999.
Phase zero is defined by the spectroscopic ephemeris at
BJD $(\phi = 0) = 2451247.50425(27) + E \times 0.11592364(43)$}
\label{t:vrad}
\begin{tabular}{cccc}
Line & Date & $K$ (km/s) & $\phi_0$ \\
\hline
\na1 (TWIN)       & March 9--11 & $286 \pm 4$ & $0.0$! \\
H$\alpha$ (TWIN)  & March 9--11 & $330 \pm 5$ & $0.008 \pm 0.002$ \\
H$\alpha$ (MOSCA) & March 18/19 & $268 \pm 8$ & $0.005 \pm 0.004$ \\
\hline
\end{tabular}
\end{table}

We could not discover any significant variation of the width of the
\na1-lines. There is a tendency of a 20\% increase of the \na1-line flux at
phase 0.5 with respect to phase 0.0. However, the flux measurements of the
\na1-lines are rather uncertain due to a too low signal-to-noise ratio of the
individual spectra and we didn't analyse this line property in more detail.

The tight phase relation between the H$\alpha$ and the \na1 lines suggests a
common origin on the secondary star. This is supported by the absence of
any further emission line component in H$\alpha$ which could be
attributed to an accretion stream or disk. The H$\alpha$ emission line is
assigned completely to the photospheric activity of the secondary star.

We regard the radial velocity
amplitude $K$ of the \na1 lines as proper measurements of the projected
orbital velocity of the secondary. Irradiation effects, which distort the
radial velocity curves in high-accretion rate polars, play an insignificant
role in \wx. The observed velocity amplitudes of H$\alpha$ in the TWIN and MOSCA
spectra are inconsistent with each other and with \na1. Hence, the centers of 
activity on the secondary were located away from the center of mass and were 
time variable on a scale of a few days.

The observed radial velocity amplitude gives some constraints on the mass
ratio and the mass of the white dwarf (see Fig.~\ref{f:qK2}). The absence of
an eclipse constrains the orbital inclination to $i < 72\degr$ assuming a
Roche-lobe filling secondary. For a Roche-lobe underfilling star the inclination
could be even higher. With the ZAMS assumption for the secondary and
the corresponding masses as listed in Tab.~\ref{t:m2par} the minimum mass of
the white dwarf is $0.36\,M_\odot$.

\begin{figure}[t]
\resizebox{\hsize}{!}{\includegraphics[clip=]{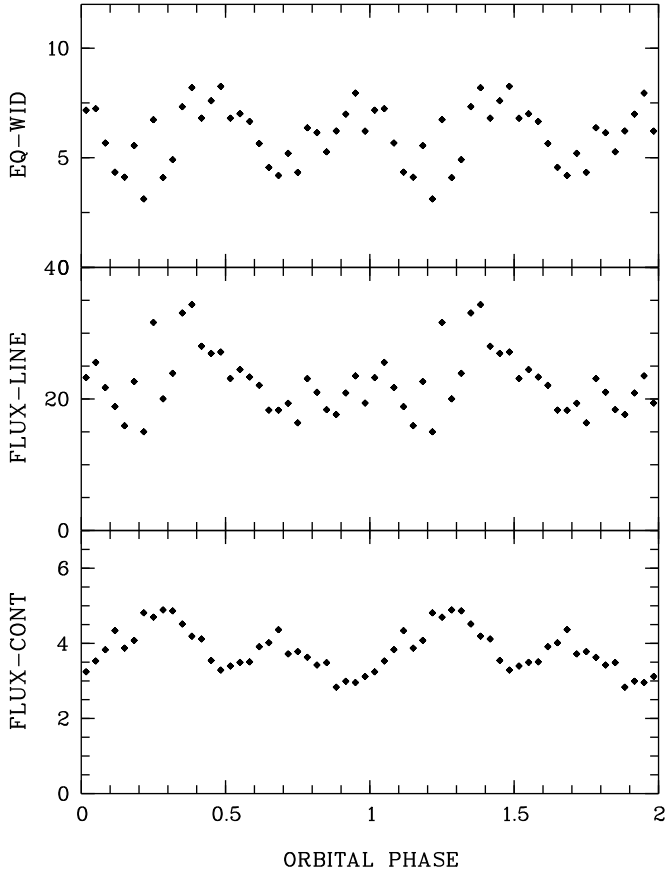}}
\caption{Phase-dependent H$\alpha$ emission line and continuum flux as well as the equivalent width as
  measured in the TWIN spectra.}
\label{f:halpha}
\end{figure}

\subsubsection{Emission line fluxes as activity indicators}
The H$\alpha$ and Ca H\&K emission lines are regarded as originating from the
active photosphere and chromosphere of the secondary. We arranged our observed
spectra like a trailed spectrogram by phase-averaging the orginal spectra into
30 phase bins. We then shifted the spectra to radial velocity zero and
subtracted the scaled spectrum of the single, non-active M4.5 star Gl 268.
The remaining flux in the spectral region around H$\alpha$ consisted of the
smoothly varying continuum from the white dwarf and the line flux without
Doppler shift and thus allowed a straightforward measurement of the line
flux. The H$\alpha$ emission line flux was related to the continuum of the
scaled Gliese star at this wavelength in order to determine the equivalent
width of the line. The results of this procedure are displayed in
Fig.~\ref{f:halpha}. 

The continuum emission light curve shows a double-humped structure with
minima at phase 0 and 0.5, respectively, very reminiscent of ellipsoidal light
variation. The $R$-band light curve of the scaled template spectrum shows
modulations of $0.2-0.3$\,mag. Using {\tt nightfall}\footnote{{\tt http://www.hs.uni-hamburg.de/DE/Ins/Per/Wichmann/ Nightfall.html}} 
we made some attempts to model this pattern. At a temperature of 3300\,K for Sp M4 one
needs a Roche-lobe filling secondary and a large inclination angle of order
70\degr~in order to get such a big amplitude. 
However, the photometric accuracy achieved by us does not
allow to draw firm conclusions on the filling factor of the secondary,
on the temperature or the inclination. E.g.~the maxima of the continuum light
curve are differing by $\sim$0.15 mag, probably a leftover from our 
calibration uncertainties. 

The line flux
itself does not show a clear phase-dependent variability pattern, 
whereas the eqivalent width again shows a pronounced
doubled-humped structure with maxima at phase 0 and 0.5, respectively. The $EW$
variied between 4 and 8\,\AA. The subtraction of the template M-star spectrum
had some uncertainty, using a possible different scaling raised $EW$ up
to 10\,\AA. These numbers were clearly smaller during the MOSCA observations
with a variation through the binary orbit between 1 and 4\,\AA.

\cite{2004PASP..116.1105W} investigated the strength of activity
via \hal\ measurements for a large number of SDSS stars. They parametrise the
strength with a $\chi$ factor, $\chi \times EW_{\rm H_\alpha} = L_{\rm
  H_\alpha}/L_{\rm bol}$. For spectral types M4 -- M4.5 the $\chi$ factor is
$\log\chi = -4.2 \dots -4.4$ and $\log(L_{\rm H_\alpha}/L_{\rm bol})$ becomes
$-3.7\dots-3.2$ for the TWIN observations. 

Recently, \cite{2004AJ....128..426W} and \cite{2005AJ....129.2428S} studied 
large number of either single \citep{2004AJ....128..426W} or common proper motion WD/MS
binaries (CPMBs, \cite{2005AJ....129.2428S}) drawn from the SDSS. 
The large majority of 90\% of the \cite{2005AJ....129.2428S}~CPMBs have an 
$EW$ close to 0 (their Fig.~2). The mean \hal\ luminosity
of the West et al.~sample at spectral type M4--M5 is $-3.57\dots-3.67$ (their
Fig.~4 and Tab.~1). These figures clearly put the
secondary in \wx\ among the more active stars in its category.
The comparison between our MOSCA and TWIN observations showed the \hal\
activity remarkably variable on a timescale of a month or shorter. 

The {\sc Ca ii} H\&K lines appear as distinct features only in the
radial-velocity corrected average spectrum of \wx\ (Fig.~\ref{f:means}). 
They can only hardly be
recognised in the original spectra where they are immersed in the noise. 
A continuum under the emission lines was determined interactively on a
graphics screen and the excess flux regarded as {\sc Ca ii} H\&K emission line
flux (with a negligible contamination from hydrogen H$\epsilon$). This gave an
integrated flux of $F_{\rm Ca} \simeq (1.5 - 2.0) \times
10^{-15}$\,erg\,cm$^{-2}$\,s$^{-1}$ and a luminosity of $\log L_{\rm Ca}
({\mbox{erg/s}}) \sim 27.3$ for an assumed distance of 100\,pc. \cite{1997A&A...325.1115P} 
have studied the relation between coronal and chromospheric
emission from cool stars in near-simultaneous ROSAT all-sky survey and Mount
Wilson data and found a good correlation between these two activity
indicators. Would the secondary follow this relation (their Fig.~4), we would
expect an X-ray luminosity of $\log L_{\rm X} ({\mbox erg/s}) \sim 33$, 
far higher than observed (see below). 

Vice versa, the observed X-ray lumonisity of \wx\
implies a {\sc Ca ii} luminosity 4 orders of magnitude less than observed,
would the star follow the Piters et al.~relation.
These estimates and comparisons suggest, that either the $L_X-L_{\rm Ca}$ 
relation between coronal and chromospheric luminosities does not apply to
the case of the fastly rotating M star in \wx\ or a very high variability of
the activity parameters of \wx\ or a combination of both. The implied X-ray
luminosity based on the observed {\sc Ca ii} luminosity would be higher than  
the typical accretion-induced luminosity in a high accretion polar. Probably
we have encountered \wx\ in a state of unusual high chromospheric activity
during the MOSCA observations. 
\begin{figure}
\resizebox{\hsize}{!}{
\includegraphics[ angle=-90, clip=]{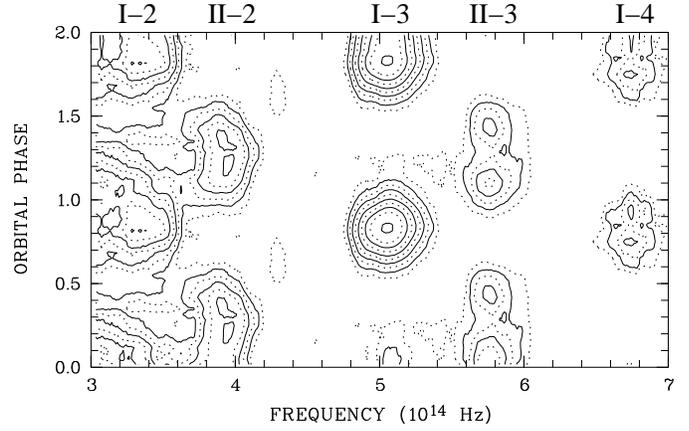}}
\caption{Contour plot of the phase-dependent cyclotron spectra from the two
  accretion regions on the white dwarf in \wx. Frequency is given in units of
  $10^{14}$\,Hz. Numerals at the top indicate frequencies for the different
  cyclotron harmonics. Roman numerals indicate primary and secondary spot, Arabic 
  numerals indicate harmonic numbers.}\label{f:trail_cyc}
\end{figure}

\subsection{Optical cyclotron radiation}
\label{cyclo}

The cyclotron spectra from the two accretion regions on the white dwarf in
\wx\ were extracted from the original data in a similar manner as described by
\cite{1999A&A...343..157R}. We firstly corrected the spectra to radial velocity
zero, arranged the spectra as a trailed spectrogram by phase-averaging the
spectra into 30 phase bins of same length 
(this left 4 bins unpopulated in the MOSCA data),
and subtracted suitably scaled spectra of our template Gl 268 (M4.5). This
gave us spectra containing mainly photospheric radiation from
the white dwarf and cyclotron radiation from the accretion spots 
(see Fig.~\ref{f:sedzoom} for a graph of the spectral decomposition). In the
following we are using the terms `accretion region' and `pole' synonymously,
just for brevity since we know about the differences.

Assuming a flat light curve from the
rotating white dwarf, a mean white dwarf spectrum was composed from spectral
regions free of cyclotron radiation and subtracted from all phase-binned
spectra. Those difference spectra are regarded as of pure cyclotron origin.
They were arranged as a trailed spectrogram and are shown on a logarithmic
intensity scale in Fig.~\ref{f:trail_cyc}. Actually, the described procedure
was performed for the red and blue TWIN spectra and the MOSCA separately and
the trailed spectrogram shown in Fig.~\ref{f:trail_cyc} is a composite of
those. The MOSCA data are used for the blue end of the spectra, $\nu > 6.3
\times 10^{14}$\,Hz, since their signal-to-noise ratio was superior to the
TWIN spectra in that wavelength region.
The feature at $\nu = 4.3 \times 10^{14}$\,Hz ($\fispe \sim 0.65$) is a
leftover from the M star subtraction. 

\begin{figure}[t]
\resizebox{\hsize}{!}{
\includegraphics[angle=-90,clip=]{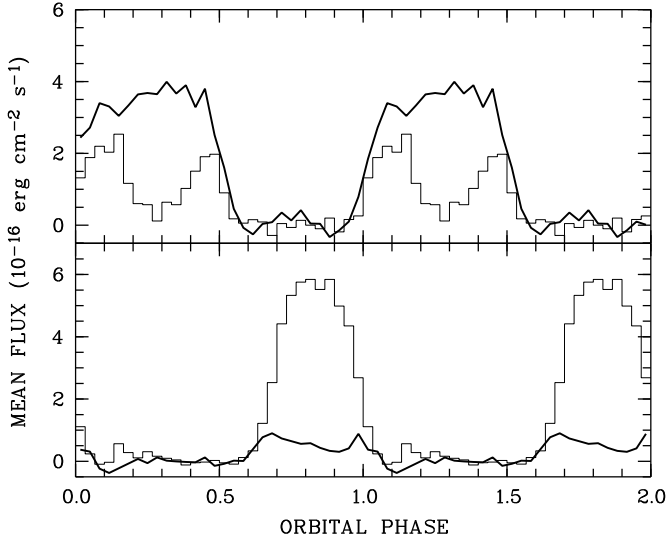}}
\caption{Light curves of the \third\ and \fourth\ (thick line) cyclotron harmonics from the primary
(bottom panel) and the \snd\ (thick line) and \third\ harmonics from the secondary (top panel)
accretion spot in \wx. The light curve for the \snd\ harmonic of the second spot is smoothed to reduce
the noise level.}
\label{f:lc_cyc}
\end{figure}

Fig.~\ref{f:trail_cyc} may be compared with the corresponding Fig.~3 of
\cite{1999A&A...343..157R}. As a novel feature we discover the $4^{\rm th}$
cyclotron harmonic from the primary pole at $\nu = 6.72 \times
10^{14}$\,Hz in the MOSCA data. All other cyclotron harmonics appear very
similar in strength, shape and position to those shown by \cite{1999A&A...343..157R}. 
Light curves for individual harmonics were computed by averaging
the trailed spectrogram over certain wavelength ranges. The light curves for
the $3^{\rm rd}$ and $4^{\rm th}$ harmonics from the brighter primary and the 
$2^{\rm nd}$ and $3^{\rm rd}$ harmonics from the fainter secondary pole are
shown in Fig.~\ref{f:lc_cyc}. Since the continuum subtraction for the lowest
frequencies was not completely satisfying, the light curve for the \snd\ harmonic of
the primary pole is ignored for the detailed analysis. 
The light curves from both poles are double-humped for the higher harmonics, an 
effect of cyclotron beaming in an optically thin plasma. The light curves of the 
lower harmonics are single-humped with maximum brightness centered between the 
two beamed humps
from the higher harmonics, indicating considerable optical depth in those
harmonics. From the centers of the bright phases, $\phi_{\rm c1} \simeq 0.84$ 
and $\phi_{\rm c2} \simeq 0.27$, the azimuth of both accretion spots were
determined to $\psi_1 \simeq 55\degr \dots 60\degr$ and  $\psi_2 \simeq
-90\degr \dots -100\degr$. 

We modeled the phase-dependent spectra
assuming homogeneous, isothermal conditions in the plasma
\citep{1981ApJ...244..569C}. A number of parameters determine
the model spectra: the field strength $B$, the temperature $kT$,
the plasma density coded with the density parameter $\log \Lambda$,
the orbital inclination $i$, the co-latitude of the field in the
accretion spot $\beta_f$, and the azimuth $\psi_f$ of the field in the
spot. The visibility of an emission region as a function of
phase is further determined by the latitude $\beta_s$ and azimuth $\psi_s$
of the accretion spot and the vertical extent of the emission
region. For simplicity, we assumed the field in the spot being normal to the
surface of the star, i.e. we set $\beta_f = \beta_s$  and $\psi_f = \psi_s$.
We further neglected any vertical extent but allowed the point-like emission
region raised to some height $h$ above the star's surface. The orbital
inclination was fixed at 70\degr.  The difference in azimuth between the
two regions was determined from the bright phase centers to $150\degr \dots
160\degr$. 

\begin{figure}[t]
\resizebox{\hsize}{!}{
\includegraphics[clip=]{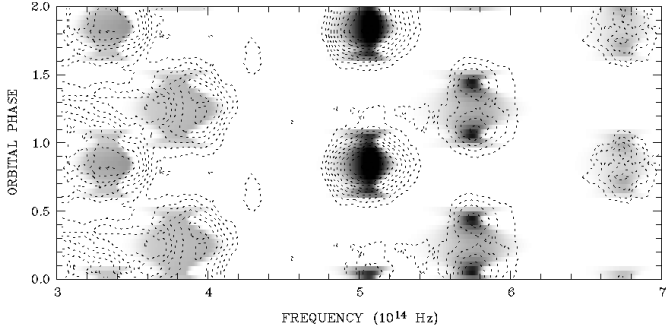}}
\caption{Comparison of the phase-dependent cyclotron model (background image)
  with the observed data (contour lines, the same as in
  Fig.~\ref{f:trail_cyc})}
\label{f:modcyc}
\end{figure}

With these simplifications, we cannot
expect to reach a perfect fit, particularly not to the light curves of
individual harmonics which are shaped by the geometry of the accretion region,
the viewing geometry, optical depth and cyclotron beaming. 
However, the main features of both cyclotron line
systems as a function of phase could be reproduced (see Fig.~\ref{f:modcyc})
by our model with the following set of parameters: 
$B_{1,2} = 61.4/69.6$\,MG, $\log\Lambda_{1,2} = 2/2$, $kT_{1,2}
 < 3$\,keV, $h_{1,2} = 0.05/0.1$\,\rwd, $\beta_{1,2} = 145\degr/135\degr$
with typical uncertainties of $\Delta B =0.5$\,MG, $\Delta\log\Lambda = 0.5$, 
$\Delta kT = 1$\,keV, $\Delta h = 0.05$\,\rwd, and  $\Delta \beta
=15\degr$. Some of the parameters are coupled to others, e.g.~if one
raises the temperature one has to lower the density parameter in order to match
the observed turnover from optically thick to thin radiation. The values given
here are slightly different from those given in \cite{2002ASPC..261..102S} based on
an analysis of the Reimers et al.~data. The differences reflect mainly the
modeling uncertainties and not a difference in the data. These were found to
be consistent with each other. The new data provide as addiditonal constraint
the beamed fourth harmonic from the second pole. 

In our modeling, the co-latitude $\beta$ was chosen according to the observed
phase separation, $\Delta \phi_b$ of the two beamed humps in either the
$4^{\rm th}$ (prime pole) or the $3^{\rm rd}$ harmonic (secondary pole). 
The value of $\beta$ is related to the orbital inclination $i$ and the length of
the self-eclipse of an accretion spot, $\Delta \phi_s = 1 - \Delta \phi_b$,
via  $\cos(\pi\Delta\phi_s) = - \cot(i) \cot(\beta)$. 
With the values of $\beta$ and $i$ fixed we had to assume a certain height of the 
region in order to reflect the length of the bright phase which lasts longer than 
the phase separation of the beams. The more extended visibility of the individual 
spots was used by \cite{2001A&A...374..189S} for their estimate of the accretion 
geometry. They determined the colatitude $\beta$ of both spots as $90\degr - 110\degr$
from light curve modeling for a point source on the surface on the WD. Using the
beaming proporties of the \third\ harmonic from the second pole the range of possible values for the colatitude 
of the second pole was extended up to $130\degr$, which is in agreement with our results. Since we 
are using solely the beaming properties of the cyclotron radiation for both spots our results seem
to give more reliable constraints than the assumption of simple geometrical foreshortening.

These results are unusual as far as the spot locations and the plasma
temperatures and densities are concerned. The two spots are
both located in the `southern' hemisphere of the white dwarf, a result already
mentioned by \cite{2001A&A...374..189S}. The opening angle between the two spots is
of order $160\degr$, i.e.~the two spots could be located at the two footpoints
of the same dipolar fieldline. If so, the southern latitude of both spots 
hints to an off-centered field configuration.
The slight difference between the field strengths in the two
spots seems to underline this, although the strong radial dependence of the
field strength does not allow to draw firm conclusions about the field at the
proper footpoints at zero height. One pole
lies in the same sector where most poles in high-accretion rate polars are
found (see \cite{1988MNRAS.231..597C} Fig.~2), i.e.~the sector leading the secondary in
phase. The other pole is away by $90\degr$ from the line connecting both stars
and away from the direction of a hypothetical accretion stream. 

The temperatures cannot be determined with very high accuracy. They are
constrained by the width of the observed highest harmonic. At 
temperatures higher than about 3\,keV the predicted line width becomes larger
than observed even for point-like emission regions. Since in nature there
will be a spread in $kT$ and $B$, we regard $\sim$3\,keV as strict upper limit.
Thus the temperatures found by us are among the lowest found 
in cyclotron spectra of polars (if \wx\ may be termed as such). 
They are more than an order of magnitude below
the shock jump temperature for an assumed 0.6\,\msun\ white dwarf.
The low temperatures together with the low density parameters $\Lambda$
imply that both cyclotron emission regions are dominated by cyclotron
cooling. It seems unlikely, that an accretion shock exists in \wx, the
accretion spots are more likely heated by particle bombardment.
The bombardment models by \cite{2001A&A...373..211F} 
relate the accretion rate per unit area, $\dot{m}$, to the field strength and 
the maximum electron temperature. According to their Fig.~5, one needs for an
0.6\,\msun white dwarf with $T_{\rm max}/T_{\rm shock} \sim 0.1$ and $B =
65$\,MG a specific mass accretion $\dot{m} \sim 0.1$\,g\,cm$^{-2}$\,\crat in
order to be deeply in the bombardement regime. 

The mean integrated cyclotron flux calculated from the observable parts of the
cyclotron spectra are $F_1 \simeq 1 \times 10^{-12}$\,\ftot\ 
and $F_2 \simeq 0.4
\times 10^{-12}$\,\ftot, respectively. The correcting factors $\kappa$ for the
unobserved parts of the cyclotron spectrum and for the beaming pattern are
rather uncertain, we assume $\kappa = 1 - 2$. We further 
assume, that the cyclotron luminosity can be calculcated as $L_{\rm cyc} =
\kappa \pi F d^2$ and get for the sum of both poles $L_{\rm cyc} \simeq (0.4 -
1) \times 10^{30}$\,erg\crat $(d/100{\mbox pc})^2$. 
The implied mass accretion rate of $\dot{M} \simeq 1.5 \times
10^{-13}$\,\mrat\, is extremely low and is of the same order as the
wind mass loss rate of the secondary star. 

Given the total and the specific mass accretion rates, one may ask for the
minimum area over which accretion may happen in order to be in the
bombardement regime. Equating $\dot{m} = \dot{M}/f_{\rm accr} 4\pi R_{\rm wd}^2$
and using $\dot{m} = 0.1$\,g\,cm$^{-2}$\,\crat, $\dot{M} \simeq 1.5 \times
10^{-13}$\,\mrat, and $R_{\rm wd} = 8\times 10^8$\,cm we obtain for two
equally sized circular accretion spots a radius of just 40\,km or $f_{\rm
  accr} \sim 1.2 \times 10^{-5}$.

\begin{figure}[t]
\resizebox{\hsize}{!}{
\includegraphics [bb = 36 36 445 701,clip=]{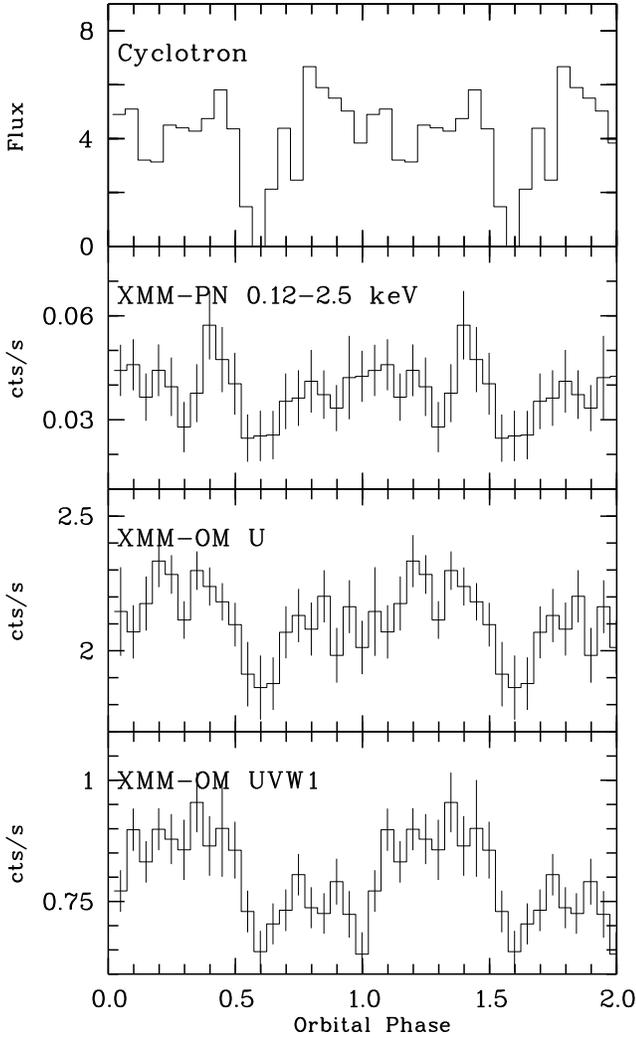}}
\caption{Phase folded light curves obtained with EPIC-PN and OM 
        together with the total cyclotron flux from Fig.~\ref{f:lc_cyc} 
        in units of 10$^{-16}$ erg cm$^{-2}$ s$^{-1}$.} 
\label{lightcurves}
\end{figure}

\begin{figure}[th]
\resizebox{\hsize}{!}{
\includegraphics[clip=]{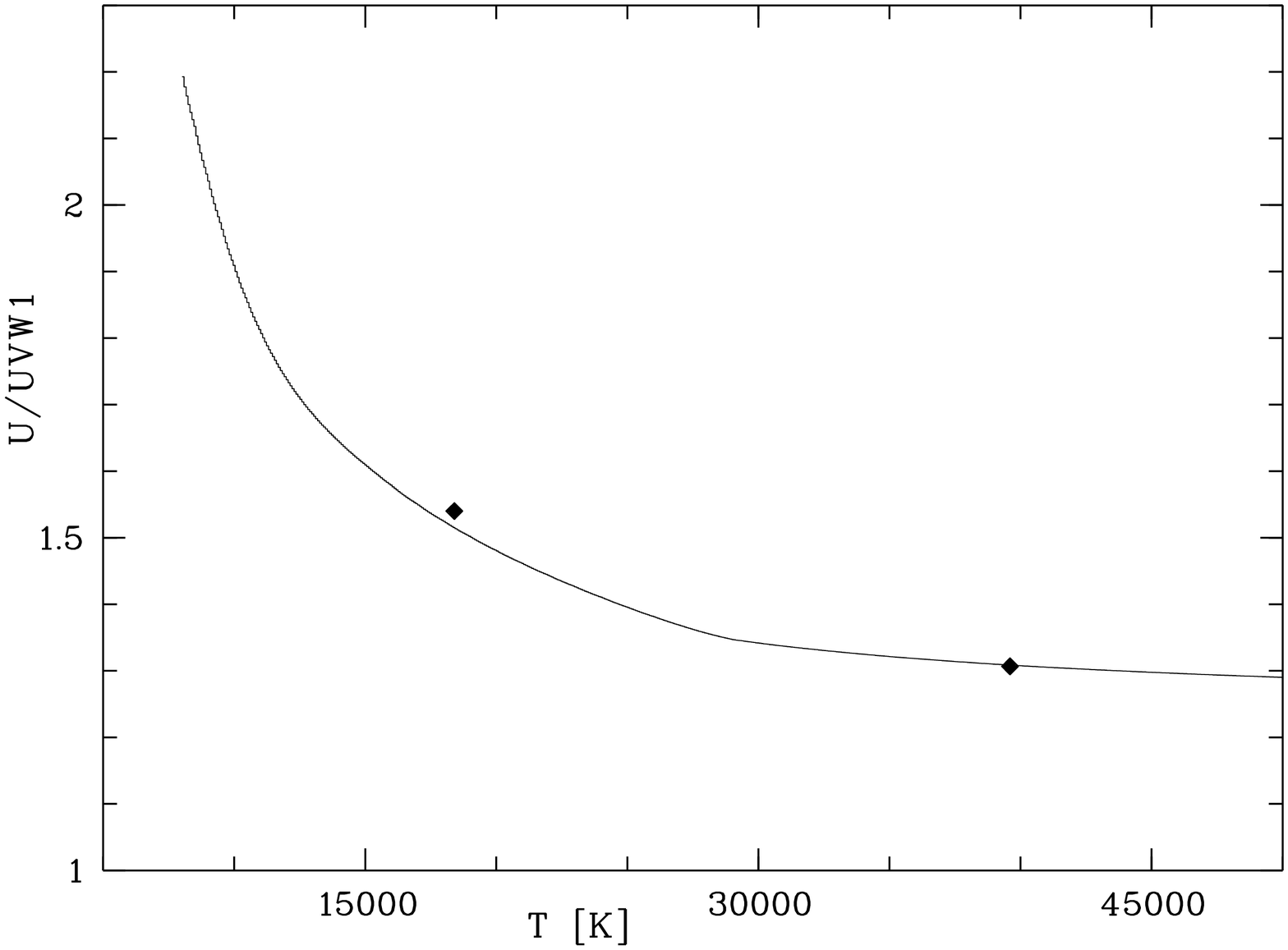}}
\caption{Theoretical count rate ratio U/UVW1 using non-magnetic model
  spectra. Overplotted are measured count rate ratios for BPM 16274 with  
  a temperature of 18700 K \citep{1995ApJ...443..735B} and GD153 with a
  temperature of 39600 K \citep{1997ApJ...480..714V}. Error bars have 
  similar sizes as the symbols.}
  \label{countrates}
\end{figure}

\subsection{XMM-Newton X-ray and ultraviolet observations}
The main results of the observations with XMM-Newton are summarized in
Figs.~\ref{lightcurves} and \ref{spec_fit}. 
Fig.~\ref{lightcurves} shows the PN X-ray light curve as well as the light
curves for both OM filters. Along with the new satellite data the summarized
flux of the \third\ and \fourth\ harmonics from the primary accretion spot and
the \snd\ and \third\ harmonics from the secondary accretion spot is shown.
The data are binned into 20 phase bins. Around phase 0.6 when both accretion spots
are self-eclipsed, the cyclotron flux drops to zero. The OM and PN light curves 
exhibit a clear variability correlated with the cyclotron flux, and thus the 
visibility of the accretion spots on the white dwarf, which is analysed in more 
detail in the following section.

\subsubsection{UV - the white dwarf atmosphere}
\label{uv}
The UV-observations with the OM were designed to estimate the temperature of
the white dwarf and of the heated photosphere below the accretion spots. The
basic assumption is that the UV light is completely dominated by photospheric
radiation from the white dwarf's surface. 
Any variability is thus assigned to temperature inhomogeneities of
the atmosphere. 
We spent each half the available observation time for the U (3440\,\AA)
and UVW1 (2910\,\AA) filters. We detected a clear, phase-dependent
variability of the OM count rate in both filters correlated with the
visibility of the accretion spots (see Fig.~\ref{lightcurves}). The ratio of
the count rates between the two filters, however, was constant within the
errors indicating only marginal T-variations.

White dwarf model spectra for pure H, $\log g = 8$, non-magnetic atmospheres
in the temperature range 8000\,K -- 100000\,K \citep{1995A&A...303..127G} were  
folded through the response curves of the OM with the different filters thus
predicting a count rate in the given filter. The count rate ratio between the
two filters as a function of temperature is shown in Fig.~\ref{countrates}. 
The ratio is a sensitive function of $\teff$ below 30000\,K. 
In order to test the reliability of our `bolometer' we retrieved calibration
observations of the well-studied white dwarfs BPM 16274 and GD 153 from the
XMM-Newton Science Archive. Their count rate ratios are also shown in the
Figure. 

The resulting count rate ratio predicts 
a temperature of $39000_{-3500}^{+4700}$\,K for GD 153 in agreement with
values found in references listed in the white dwarf data
base\footnote{http://procyon.lpl.arizona.edu/WD/}, $37900 - 40100$\,K. 
The derived temperature of $17340_{-80}^{+150}$\,K for BPM 16274 is somewhat
lower than the values of $18400 - 18700$\,K found in the literature. We thus
estimate the accuracy of the filter ratio method of order 10\%.

The observed count rate ratio around phase $\phi = 0.6$, when both spots
are hidden from the observer, is $2.88\pm 0.26$.
This high ratio cannot be reproduced by our model atmospheres. The observed ratio 
indicates a softer spectrum, i.e. cooler atmosphere than the lower limit of 
8000\,K of our models. 
Synthetic magnetic white dwarf model spectra with dipolar field structure for
a pole field strength of 60 MG -- kindly provided by S.~Jordan -- reduced
this discrepancy, but also led to no satisfying result. It need to be said,
however, that the model spectrum of a magnetised atmosphere used the
temperature structure of a non-magnetic atmosphere, hence we cannot quantify
to what extent the magnetic model matches the observations better than the
non-magnetic. 

It is worth noting at this stage, that we could not identify any depression (a
line or a trough) in the optical continuum of the observed white dwarf
spectrum, which could be even only tentatively identified as a Zeeman signal.
We guess, that the low $\teff$ and/or a complex field structure are responsible
for this observation. 

We conclude that $\teff$ of the white dwarf is almost certainly below 8000\,K.
Even without using the count rate ratio as an indicator for the temperature
the measured count rate in the UVW1 filter around phase $\phi = 0.6$ gives an 
upper limit for the flux and thus the temperature of the white dwarf.
Therefore one has to assume that the whole white dwarf is seen, without contribution 
from the heated pole caps. If we fix the distance at 100 pc as derived in section 
\ref{dist} and use a mean polar white dwarf mass of 0.6 $M_{\odot}$ 
\citep{2000PASP..112..873W}, the lower limit of 8000\,K of our nonmagnetic 
model atmospheres gives a flux which is about 12\% higher than the measured flux 
in the UVW1 filter. We use the mass-radius-relation from \cite{1972ApJ...175..417N}, 
which is sufficient for very cool white dwarfs \citep{2000A&A...353..970P}. Any 
additional contribution from the accretion spots decreases the flux contributed by 
the white dwarf and thus its temperature. A significantly higher temperature would only be 
possible for a very massive white dwarf or if it were much further away than the 
derived distance of 100 pc.

\begin{figure}[t]
\resizebox{\hsize}{!}{

\includegraphics[clip=]{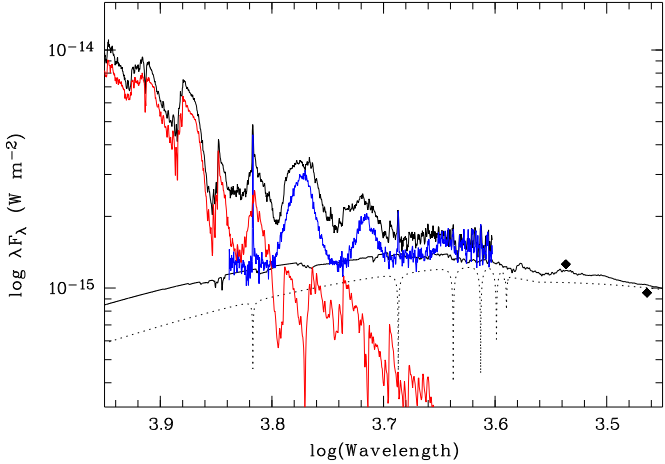}}
\caption{Observed and deconvolved optical spectrum and ultraviolet photometry
  of \wx. Shown are the observed mean spectrum (black top curve), the scaled
  spectrum of Gl268 (red), the M-star subtracted spectrum (cyclotron radiation
  plus white dwarf, blue). The rhombs denote the OM photometry and the black
  lines at bottom white dwarf model spectra for a distance of 100\,pc. 
  The black dotted line is the
  non-magnetic white dwarf model with $M_{\rm wd} = 0.6$\,\msun, the black
  solid line denotes a magnetic white dwarf model with 
  $M_{\rm wd} = 0.47$\,\msun, $B = 60$\,MG, $d = 100$\,pc.}
         \label{f:sedzoom}
\end{figure}

\begin{figure}[t]
\resizebox{\hsize}{!}{
\includegraphics[clip=]{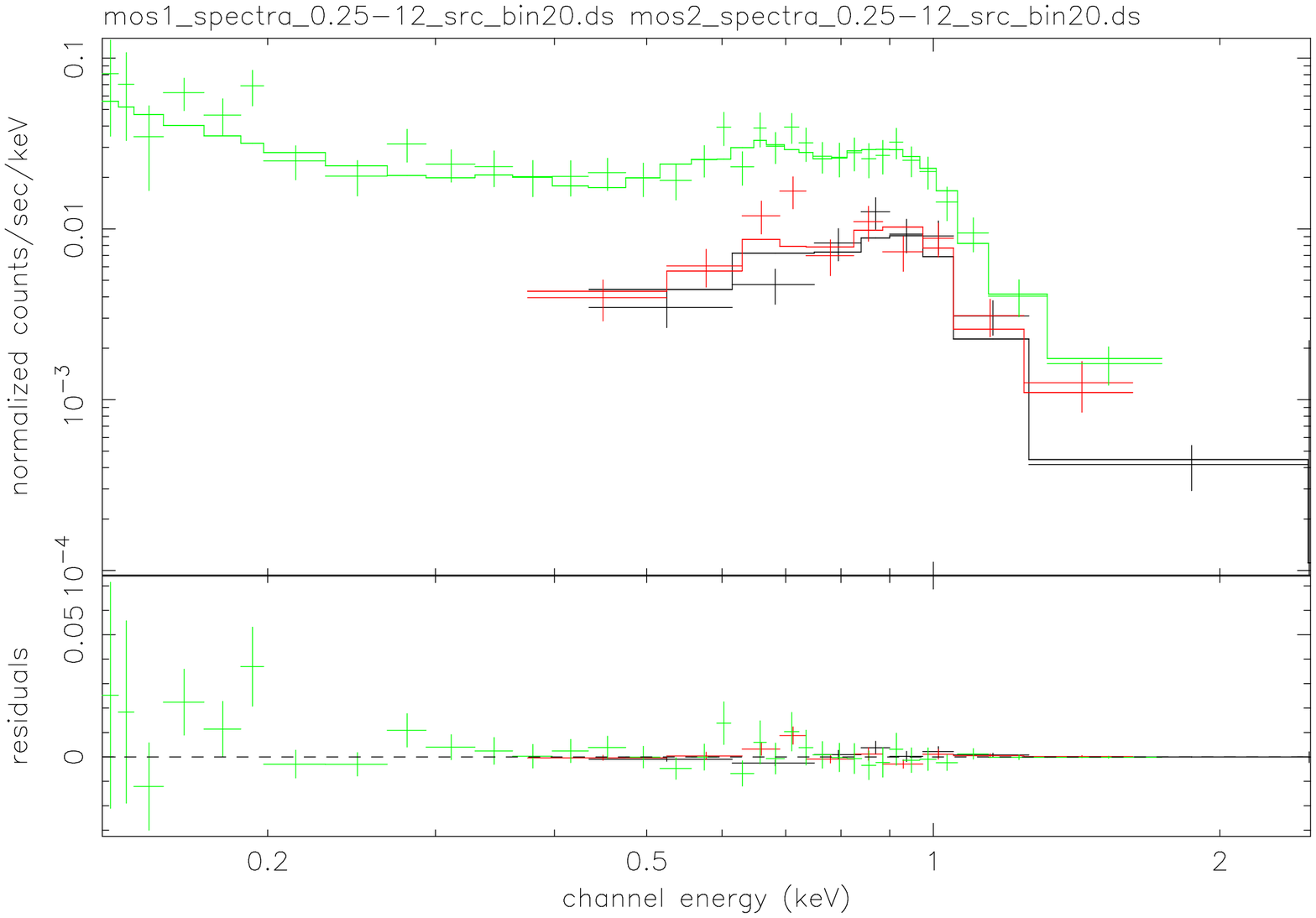}}
\caption{Spectral fit of the combined PN and MOS spectra with a two component
  MEKAL model (see Tab.~\ref{table:1} for spectral parameters)}.
\label{spec_fit}
\end{figure}

The different spectral components in the optical and near ultraviolet are
illustrated in Fig.~\ref{f:sedzoom} which shows the mean optical spectrum, the
scaled spectrum of the M4.5 template (Gl268), the blue TWIN spectrum after
subtraction of the M dwarf (white dwarf plus cyclotron radiation), the two UV
measurements and suitably scaled magnetic and non-magnetic white dwarf model
spectra for $\teff = 8000$\,K.

\cite{1999A&A...343..157R} determined the temperature of the white dwarf in
\wx\ to be $13000 \pm 1000$\,K. Our new determination differs significantly. 
Reimers et al.~substracted a dM4.5e (G3-33)  
spectrum from a mean spectrum when both spots are self-eclipsed. The spectral energy  
distribution of the remaining spectrum was then compared with synthetic
spectra for white dwarfs with a magnetic field of 60 MG.  
The temperature dependence of the continuum slope in the wavelength range 
between 4000 to 6000\,\AA~used by them is not very pronounced. 
Due to the much smaller contribution of the secondary to the total flux in 
the UV, the use of the  OM data seems more reliable to us for a determination 
of the white dwarf's temperature and we thus reject their temperature estimate.
A white dwarf with a temperature of $13000 \pm 1000$\,K would have been clearly
detected in the OM, even at a distance of 140 pc.

The unusual cool temperature of the white dwarf fits well in the current
knowledge of all LARPs. Table \ref{table:2} summarizes the white dwarf
temperatures as determined so far. Apart from SDSS0837 all white dwarfs are
cooler than 11000\,K, which otherwise seems to be a lower limit for the
white dwarfs in normal CVs \citep{2000RvMA...13..151G}. Without any
prior accretion the temperature of the white dwarf in \wx\ implies a 
cooling age in the order of 10$^9$ years \citep{2000ApJ...543..216C}
as a lower limit.
\cite{2003A&A...406..305S} deduced a typical duration of the post common envelope
phase of about 2 Gyr. Thus, CVs on the initiation of accretion due to Roche-lobe 
overflow should contain very cool white dwarfs. The low temperature also restricts the 
accretion rate in the past to have been less than 10$^{-11}$ M$_{\odot}$ yr$^{-1}$, 
at least for periods shorter than the thermal timescale of the heated envelope 
\citep{2003ApJ...596L.227T}. Both limits support the assumption that \wx\ and the
LARPs in general are pre-CVs before the onset of Roche-lobe overflow.

The UV light curves in Fig.~\ref{lightcurves} show a clear variability
correlated with the visibility of the accretion spots, a clear indication for 
heated pole caps. Surprisingly the secondary spot seems to be hotter than the
primary pole, which is inconsistent with the picture obtained from the 
optical spectra and the X-ray light curve. The reason for this is rather
unclear. We can exclude a cyclotron contamination because the  OM U-filter
is centered between harmonics 4 and 5 of the second pole.

The hardness ratio for the OM filters shows no significant variability which
is somewhat confusing since our interpretation of photometric variability of 
the UV requires also a hardness ratio variability. However, we cannot
completely exclude that those expected count rate ratio variations are below
our sensitivity threshold and cannot test this due to the lack of suitable
model atmospheres. 

\begin{table*}
\begin{minipage}[t]{\columnwidth}
\caption{Spectral fits for the combined PN and MOS spectra}             
\label{table:1}      
\centering                          
\renewcommand{\footnoterule}{}  
\begin{tabular}{lclcccc}        
\hline\hline                 
Model                    & $\chi_{\nu}^2$  & NHP$^a$      & kT (MEKAL)              & kT (BB)& n$_H^b$      & Flux$^c$\\   
                         &                 &              & [keV]                   & [eV]            & ~~~~~~~[cm$^{-2}$]  & ~~~~~~~[ergs cm$^{-2}$ s$^{-1}$]\\
\hline                                                                                                    
wabs (MEKAL)             & ~~~~~~3.44      & ~~~~1.3e-15  & 0.62$ \pm $0.02         &                 & ~~~~~~~5.6e-6       & ~~~~~~~4.2E-14\\
wabs (BB+MEKAL)          & ~~~~~~1.42      & ~~~~2.6e-2   & 0.66$ \pm $0.02         & 117 $\pm$ 11    & ~~~~~~~1.0$^d$      & ~~~~~~~4.6E-14\\
wabs (MEKAL+MEKAL)       & ~~~~~~0.98      & ~~~~0.51     & 0.26$ \pm $0.02         &                 & ~~~~~~~1.2e-12      & ~~~~~~~5.4E-14\\
                         &                 &              & 0.82$ \pm $0.05
                         &                 &                     &\\
\hline
\end{tabular}
\end{minipage}
\begin{center}
\begin{minipage}[h]{\textwidth}
\begin{flushleft}
$^a$ null hypothesis probability, $^b$ in units of 1e20, $^c$ integrated flux 0.1-5 keV, $^d$ frozen 
\end{flushleft}
\end{minipage}
\end{center}
\end{table*}

\subsection{The X-ray light curves and spectra}
The EPIC-PN light curve shows orbital phase-dependent modulation with a
pronounced minimum at the time, when both accretion spots are hidden from the
observers view. While the cyclotron flux drops to zero around $\phi \sim 0.6$, 
the X-ray flux remains at the constant minimum flux. We ascribe the residual 
flux around that phase to the corona of the secondary star. 

The average count rate of \wx\ is too small to make a phase-dependent spectral
analysis feasible. Fig.~\ref{spec_fit} shows the orbital mean spectra obtained
with EPIC-PN and EPIC-MOS. Essentially all the X-ray flux emerges below 2
keV. We fitted the  combined PN and MOS spectra with XSPEC. Our results are
summarized in Table \ref{table:1}. A first approach
with a single MEKAL model for a coronal and/or accretion plasma, molding 
emission from the secondary and from the accretion region with one
temperature led to no satisfying fit. Motivated by the presence of a soft
blackbody-like and a hard thermal component in high accretion rate 
polars, our next attempt included a black-body component.
The fit was improved, but the temperature of $117 \pm 11$\,eV for the black
body is higher than in any other polar observed so far and thus seems 
implausible. A satisfactory fit  was achieved using a two component MEKAL 
model. The temperatures of $0.26\pm 0.02$\,keV and $0.82 \pm 0.05$\,keV are
well within the regime that one  could expect for a coronal plasma. The
temperature thus determined would be very low for the accretion plasma of a
normal-accreting polar. It is however in the same range as the temperatures
derived from the analysis of the cyclotron emitting plasma. 

To make a possible distinction between the two components (accretion plasma
and corona) we extracted spectra according to the time 
intervals when only one of the two spots is directed towards the observer.  
The spectral fits of those two separated spectra yielded no significant
differences between each other and compared to the overall mean spectrum.  
Also, we could not detect any significant variation in the X-ray hardness
ratio as a function of orbital phase. Hence, we cannot discern between the
coronal and the accretion plasma on the basis of possible spectral
differences with the given number of photons.

\begin{figure}
\resizebox{\hsize}{!}{
   \includegraphics[angle=-90,clip=]{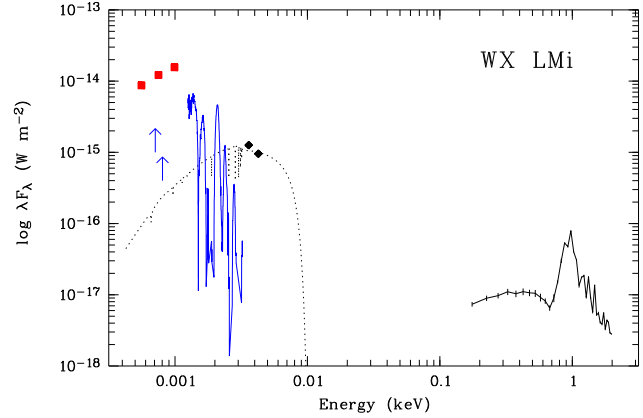}}
\caption{Spectral energy distribution for \wx\ combining photometry in JHK
   (2MASS, red squares) representing the secondary star, the OM onboard
   XMM-Newton (black rhombs) and the 8000\,K nonmagnetic white dwarf model
   representing the white dwarf (0.6 $M_{\odot}$, d=100pc), the EPIC-PN spectrum as a mix of radiation
   from the accretion and the coronal plasma, and the sum of the two cyclotron
   components (blue line). The arrows indicate the wavelengths of the cyclotron
   fundamentals of the two poles.}
\label{sed}
\end{figure}

Figure~\ref{sed} shows the spectral energy distribution from the near
infrared to the X-ray regime for the individual spectral components.
The graph shows $\lambda F_{\lambda}$, i.e.~the values plotted 
are representing the contributions to the total energy output from the 
system. 
Red squares indicate the contribution of the secondary, which dominates the
system absolutely. Black rhombs indicate the OM data and, together with the
dashed line, indicate the contribution from the white dwarf, which can 
purely be observed in the ultraviolet. 
The spectrum shown in blue is the summed cyclotron spectrum from both
poles. Arrows indicate the position of the cyclotron fundamental which are
located in the $H$-band and between the $H$ and $K$ bands, respectively,
i.e.~the $K$-band is likely uncontaminated by cyclotron radiation. 
The X-ray spectrum, shown as black solid line, is a mix of thermal plasma
radiation from the accretion poles and the coronal plasma. Accretion-induced
radiation is clearly dominated by the cyclotron component.

The unabsorbed flux (0.1 - 5 keV) for the two component MEKAL fit is $5.4
\times 10^{-14}$\,erg\,cm$^{-2}$\,s$^{-1}$. For a distance of 100\,pc this
yields an X-ray luminosity of $6.4 \times 10^{28}$\,erg\,s$^{-1}$.
Based on the variability of the X-ray light curves we assume that the secondary 
and the accretion regions contribute roughly equal X-ray flux and luminosity to 
the observed X-ray spectrum, i.e.~we get  
$L_{\rm{M_2,X}} \sim (L_{\rm{1,X}} + L_{\rm{2,X}}) \sim 3.2 \times 10^{28}$\,erg\,s$^{-1}$.

\subsubsection{The secondary in X rays}
\label{secxrays}

Our estimate of the X-ray luminosity of the secondary star makes a comparison 
with the X-ray activity of normal late-type dwarfs possible. The latter is
correlated with the stellar rotation \citep{1981ApJ...248..279P, 2003A&A...397..147P}. 
According to \cite{1996ApJS..104..117L} the bolometric luminosity of our
assumed M4.5 secondary is $L_{\rm{bol}}$ = 3.9 $\times$ 10$^{31}$ erg
s$^{-1}$. The X-ray flux attributed to the secondary,  
$L_{\rm{M_2,X}} \sim 3.2 \times 10^{28}$\,erg\,s$^{-1}$, results in 
$L_{\rm{X}}$/$L_{\rm{bol}} \simeq 0.8 \times 10^{-3}$. 
This result is only weakly dependent on the assumed spectral type, i.e.~ for
an M3 secondary we get $0.3 \times 10^{-3}$.\\
\cite{2003A&A...397..147P} studied the relation between
coronal X-ray emission and the rotation period. They confirmed the existence
of two regimes, one in which the rotation period is related to the X-ray
emission and one in which a constant saturated X-ray to bolometric luminosity
is attained (their diagrams 5 and 6). This claim for the latest spectral types
rests on rather small number of stars. The X-ray luminosity determined for the
secondary of \wx\ falls well on the constant branch of their diagrams, 
suggesting saturation of the activity.

\begin{table*}
\begin{minipage}[h]{\textwidth}
\caption{Summary of system parameters} 
\label{table:2}     
\centering                    
\renewcommand{\footnoterule}{}
\begin{tabular}{llllccccccc}       
\hline\hline
Object       &  Period  & Spectral type$$ & Comparison$^b$ & Distance$$ & $L_{bol}$  & $L_X$  & $L_X / L_{bol}$ & $T_{WD}$  & Reference$^c$\\
             &           &                 &               & [pc]       &             & [erg/s]   & $[10^{-3}]$     & [10$^3$ K]    &\\
\hline
\wx          &  2.78     & M4.5            & Gl268  & 100        & 3.86e31    & $~~$ 3.2e28$^a$& $~~$ 0.8         & $<$ ~8                 & 4\\      
HS0922       &  4.07     & M3.5            & Gl494  & 190        & 1.97e32    & $<$ 7.5e29    & $<$ 3.8         & $<$ 10                & 1,3\\
SDSS0837     &  3.18     & M5              & LHS377 & 330        & 3.43e30    & -             &  -              & $<$ 14                & 3\\
SDSS1324     &  2.6      & M6              & GJ1111 & 450        & 3.18e30    & $<$ 1.4e29    & $<$ 46          & $<$ ~6                 & 1,2,3\\
SDSS1553     &  4.39     & M5              & LHS377 & 130        & 3.43e30    & $~~$ 1.4e29    & $~~$ 41          & $\sim$ ~5.5            & 1,2,3\\
SDSS2048     &  4.2      & M3              & Gl388  & 260        & 8.93e31    & $<$ 7.9e29    & $<$ 0.9         & $<$ ~7.5               & 3\\
\hline
\end{tabular}
\end{minipage}
\begin{center}
\begin{minipage}[h]{0.9 \textwidth}
\begin{flushleft}
$^a$ taken one half of the unabsorbed MEKAL+MEKAL flux in the range 0.1 - 5 keV\\
$^b$ see \cite{1996ApJS..104..117L}\\
$^c$ References: (1)\cite{2005ASPC..330..137W}, (2)\cite{2004AJ....128.2443S}, (3)\cite{2005ApJ...630.1037S} and references therein, (4) this paper
\end{flushleft}
\end{minipage}
\end{center}
\end{table*}
The same comparison can be made for the other low accretion rate polars as
well. Table~\ref{table:2} compiles the parameters from the literature. 
Among the other systems only SDSS1553 has a clear X-ray detection and the 
source was found clearly above the Pizzolato et al.~relation. However, 
the luminosity variations for the late spectral types in the compilation 
of \cite{1996ApJS..104..117L} are large. Taking into account the uncertainty 
for the spectral type of the secondary, SDSS1553 could satisfy the Pizzolato et al.~relation
as well. Otherwise this could indicate additional X-ray emission from the accretion plasma,
although the PN light curve shows no variability, contrary to the optical
light curve \citep{2004AJ....128.2443S}. 
For SDSS1324 only an upper limit was obtained from XMM observation
\citep{2004AJ....128.2443S} and for HS0922 there is an upper limit from a
ROSAT PSPC observation \citep{2000A&A...358L..45R}. SDSS2048 could be 
identified with a RASS source, but there are other optical objects within the
error circle \citep{2005ApJ...630.1037S}, which makes the identification as 
an X-ray source ambiguous. If we take the upper limit as face value, it falls
on the relation by Pizzolato.
Hence, one (\wx) of the two objects with a proper X-ray detection seems to behave like
a normal main-sequence star as far as X-ray activity is concerned and lends
support to the picture of saturated activity, the other leads to no definite conclusion,
concerning the uncertainty in L$_{\rm{bol}}$.

\section{Main results and discussion}
\label{conclu}
We have presented a photometric and spectroscopic analysis at optical,
ultraviolet and X-ray wavelengths of \wx, a system termed in the past
as a low-accretion rate polar (LARP). This designation was chosen due 
to its similarity to the polars, normal accreting magnetic CVs, as far as its
stellar constituents and the magnetic field strength are concerned. 

Our main results can be summarized as such: \wx\ harbours a highly active
secondary star of spectral type dM4.5 $\pm0.5$. The acitvity is obvious from
photospheric \hal\ and  chromospheric {\sc Ca ii} lines, which vary in phase
with the orbital motion of the secondary, although not centered on the stellar
disk. Activity is detected also at X-ray wavelengths at a level comparable to
other late-type stars at saturation. 

The phasing of the \hal-lines sheds some new light on old observations
of polars in their low states. Faint residual \hal emission moving in phase
with the secondary (traced with the \Na lines) was observed e.g.~in UZ
For \citep{1997A&A...320..181S} or MR Ser \citep{1993A&A...278..487S}, although typically
at a different velocity than the \Na lines. While \hal-emission was
typically interpreted in terms of irradiation, it may be that those low state
polars revealed their active secondaries. Low-state spectrophotometry 
of polars thus might further constrain the activity-rotation relation in a
rather unexplored regime. 
This active secondary could be an explanation for the missing magnetic white 
dwarf + main-sequence binaries in current catalogs \citep{2005AJ....129.2376L}. 
If the magnetic white dwarf captures the wind from the active secondary, the emission
of cyclotron radiation in the optical clearly vitiates the color selection criteria and/or
the spectral composition. It requires a strongly magnetic white dwarf ($B >$ 50 MG) to
efficiently capture the wind, but the LARPs could mitigate the lack of those binaries
at least for the case of highly magnetic white dwarfs.

At the given orbital period, $\porb = 2.78$\,h, one would according to
\cite{2000NewAR..44...93B} expect a secondary of spectral type M3 rather than M4.5. In
this respect the secondary seems to behave as those in normal accreting polars
which also are observed typically 'too cool to comfort' \citep{1990MNRAS.246..637F}. 
However, the standard explanation assuming the secondaries being
driven out of thermal equilibrium does not seem to work here, since the low 
temperature of the white dwarf, $\teff < 8000$\,K, implies the absence of 
or a very
low accretion rate in the past. If \wx\ and its relatives would be just in an
occasional low state of accretion the expected temperature should be above
11000\,K \citep{2000RvMA...13..151G}, as for the white dwarfs in  normal
CVs. The observed accretion rate is too small for a long-term
equilibrium mass transfer  \citep{2005ASPC..330..137W}. 
Also, the
location of the second pole in \wx\ makes accretion via Roche-lobe overflow
rather unlikely. 

As suggested already by \cite{2001A&A...374..189S} for \wx\ and  \cite{2005ApJ...630.1037S}
for similar systems found in the SDSS the secondary is likely 
somewhat underfilling. 
Thus our analysis of \wx\ fully supports the view of LARPs as pre-Polars,
post-common-envelope systems prior to Roche lobe contact, like already
concluded  by \cite{2005ASPC..330..137W} and \cite{2005ApJ...630.1037S}.
We nevertheless detected ellipsoidal light modulations
which in principle could be used to constrain the inclination and the filling
factor, provided spectrophotometric data with higher accuracy could be obtained.

We re-determine the distance to be $d = 100\pm20$\,pc. The accretion
luminosity, which is dominated by optical cyclotron radiation with a minor
contribution from X-ray thermal plasma emission, implies a low mass transfer
rate of $\dot{M} \simeq 10^{-13}$\,\mrat. Such low mass transfer rates are
likely compatible with the wind mass loss rate from the active
secondary. \cite{2005ApJ...630.1037S}, adopting a scenario by \cite{1995MNRAS.276..255L}, 
proposed a `magnetic siphon' channeling all material in the wind down to the
pole caps with very little wind loss. The actual accretion geometry, determined
by us from the orbital phases of the cyclotron features, doesn't seem to play
an important role, once the magnetic field is sufficiently strong. 

\cite{2001A&A...374..189S}, modeling optical light curves and taking the beaming
properties for the second spot into account, located the spots near
the equator. We refine this model by a phase-dependent study of the cyclotron
beaming for both spots. Primary and secondary accretion spot are located on the `southern' hemisphere, i.e.~away from
the observer's hemisphere, indicating a field structure different from a
centered dipole. The field strengths in the two spots are 61.4\,MG and
69.6\,MG, respectively.

At such high field strength and low accretion rates as predominating in LARPs
probably no accretion shock forms. As shown by our deconvolution of the
spectral energy distribution for the infared to the X-ray regime, 
the plasma cools mainly via cyclotron radiation instead of bremsstrahlung as
in high-accretion rate polars lending full support to the bombardement
solution by \cite{2001A&A...373..211F}.

\wx\ as the brightest of the new class of pre-polars is a suitable target for
spectropolarimetric observations at a 10m class telescope in order to
search for polarized signal from the white dwarf's photosphere and to perform
detailed modeling of the accretion plasma on its poles.

\begin{acknowledgements}
We thank S.~Jordan for providing a grid of model spectra of magnetic white
dwarfs. We also thank our anonymous referee for helpful comments and for 
pointing some inconsistent wording in the original version of the paper.
JV is supported by the Deutsches Zentrum f\"ur Luft- und Raumfahrt
(DLR) GmbH under contract No. FKZ 50 OR 0404. BTG was supported by a PPARC Advanced Fellowship.
\end{acknowledgements} 

\bibliographystyle{aa}
\bibliography{references}
\end{document}